# Perfect Matchings via Uniform Sampling in Regular Bipartite Graphs

Ashish Goel[*]    Michael Kapralov [†]    Sanjeev Khanna[‡]

September 24, 2018

## Abstract

In this paper we further investigate the well-studied problem of finding a perfect matching in a regular bipartite graph. The first non-trivial algorithm, with running time $O(mn)$, dates back to König's work in 1916 (here $m = nd$ is the number of edges in the graph, $2n$ is the number of vertices, and $d$ is the degree of each node). The currently most efficient algorithm takes time $O(m)$, and is due to Cole, Ost, and Schirra. We improve this running time to $O(\min\{m, \frac{n^{2.5} \ln n}{d}\})$; this minimum can never be larger than $O(n^{1.75}\sqrt{\ln n})$. We obtain this improvement by proving a uniform sampling theorem: if we sample each edge in a $d$-regular bipartite graph independently with a probability $p = O(\frac{n \ln n}{d^2})$ then the resulting graph has a perfect matching with high probability. The proof involves a decomposition of the graph into pieces which are guaranteed to have many perfect matchings but do not have any small cuts. We then establish a correspondence between potential witnesses to non-existence of a matching (after sampling) in any piece and cuts of comparable size in that same piece. Karger's sampling theorem for preserving cuts in a graph can now be adapted to prove our uniform sampling theorem for preserving perfect matchings. Using the $O(m\sqrt{n})$ algorithm (due to Hopcroft and Karp) for finding maximum matchings in bipartite graphs on the sampled graph then yields the stated running time. We also provide an infinite family of instances to show that our uniform sampling result is tight up to poly-logarithmic factors (in fact, up to $\ln^2 n$).

## 1 Introduction

A bipartite graph $G = (U, V, E)$ with vertex set $U \cup V$ and edge set $E \subseteq U \times V$ is said to be regular if every vertex has the same degree $d$. We use $m = nd$ to denote the number of edges in $G$ and $n$ to represent the number of vertices in $U$ (as a consequence of regularity, $U$ and $V$ have the same size). Regular bipartite graphs have been the subject of much study. Random regular bipartite graphs represent some of the simplest examples of expander graphs [12]. These graphs are also used to model scheduling, routing in switch fabrics, and task-assignment problems (sometimes via edge coloring, as described below) [1, 6].

A regular bipartite graph of degree $d$ can be decomposed into exactly $d$ perfect matchings, a fact that is an easy consequence of Hall's theorem [4]. Finding a matching in a regular bipartite graph is a well-studied problem, starting with the algorithm of König in 1916, which is now known to run in time $O(mn)$ [11]. The well-known bipartite matching algorithm of Hopcroft and Karp [8] can be used to obtain a running time of $O(m\sqrt{n})$. In graphs where $d$ is a power of 2, the following simple idea, due to Gabow and Kariv [7], leads to an algorithm with $O(m)$ running time. First, compute an Euler tour of the graph (in time $O(m)$) and then follow this tour in an arbitrary direction. Exactly half the edges will go from left to right; these form a regular bipartite graph of degree $d/2$. The total running time $T(m)$ thus follows the recurrence $T(m) = O(m) + T(m/2)$ which yields $T(m) = O(m)$. Extending this idea to the general case proved quite hard, and after a series of improvements (eg. by Cole and Hopcroft [5], and then by Schrijver [13] to $O(md)$), Cole, Ost, and Schirra [6] gave an $O(m)$ algorithm for the case of general $d$.

[*]Departments of Management Science and Engineering and (by courtesy) Computer Science, Stanford University. Email: ashishg@stanford.edu. Research supported by NSF ITR grant 0428868 and NSF CAREER award 0339262.
[†]Institute for Computational and Mathematical Engineering, Stanford University. Email: kapralov@stanford.edu.
[‡]Department of Computer and Information Science, University of Pennsylvania, Philadelphia PA. Email: sanjeev@cis.upenn.edu. Supported in part by a Guggenheim Fellowship, an IBM Faculty Award, and by NSF Award CCF-0635084.

The main interest of Cole, Ost, and Schirra was in edge coloring of general bipartite graphs of maximum degree $d$, where finding perfect matchings in regular bipartite graphs is an important subroutine. Finding perfect matchings in regular bipartite graphs is also closely related to the problem of finding a Birkhoff von Neumann decomposition of a doubly stochastic matrix [3, 16].

In this paper we present an algorithm for finding a perfect matching in a regular bipartite graph that runs in time $O(\min\{m, \frac{n^{2.5}\ln n}{d}\})$. It is easy to see that this minimum can never be larger than $O(n^{1.75}\sqrt{\ln n})$. This is a significant improvement over the running time of Cole, Ost, and Schirra when the bipartite graph is relatively dense. We first prove (Theorem 2.1 in section 2) that if we sample the edges of a regular bipartite graph independently and uniformly at rate $p = O(\frac{n \ln n}{d^2})$, then the resulting graph has a perfect matching with high probability. The resulting graph has $O(mp)$ edges in expectation, and running the bipartite matching algorithm of Hopcroft and Karp gives an expected running time of $O(\frac{n^{2.5}\ln n}{d})$. Since we know this running time in advance, we can choose the better of $m$ and $\frac{n^{2.5}\ln n}{d}$ in advance. It is worth noting that uniform sampling can easily be implemented in $O(1)$ time per sampled edge assuming that the data is given in adjacency list format, with each list stored in an array, and assuming that $\log n$ bit random numbers can be generated in one time step[1].

We believe that our sampling result is also independently interesting as a combinatorial fact. The proof of our sampling theorem relies on a sequential decomposition procedure that creates a vertex-disjoint collection of subgraphs, each subgraph containing many perfect matchings on its underlying vertex set. We then show that if we uniformly sample edges in each decomposed subgraph at a suitably chosen rate, with high probability at least one perfect matching survives in each decomposed subgraph. This is established by using Karger's sampling theorem [9, 10] in each subgraph. An effective use of Karger's sampling theorem requires the min-cuts to be large, a property that is not necessarily true in the original graph. For instance, $G$ could be a union of two disjoint $d$-regular bipartite graphs, in which case the min-cut is 0; non-pathological examples are also easy to obtain. However, our serial decomposition procedure ensures that the min-cuts are large in each decomposed subgraph. We then establish a 1-1 correspondence between possible Hall's theorem counter-examples in each subgraph and cuts of comparable size in that subgraph. Since Karger's sampling theorem is based on counting cuts of a certain size, this coupling allows us to claim (with high probability) that no possible counter-example to Hall's theorem exists in the sampled graph. On a related note, Benczur [2] presented another sampling algorithm which generates $O(n \ln n)$ edges that approximate *all cuts*; however this sampling algorithm, as well as recent improvements [15, 14] take $\tilde{\Omega}(m)$ time to generate the sampled graph. Hence these approaches do not directly help in improving upon the already known $O(m)$ running time for finding perfect matchings in $d$-regular bipartite graphs.

---
[1]Even if we assume that only one random bit can be generated in one time step, the running time of our algorithm remains unaltered since the Hopcroft-Karp algorithm incurs an overhead of $\sqrt{n}$ per sampled edge anyway.

The sampling rate we provide may seem counter-intuitive; a superficial analogy with Karger's sampling theorem or Benczur's work might suggest that sampling a total of $O(n \ln n)$ edges should suffice. We show (Theorem 4.1, section 4) that this is not the case. In particular, we present a family of graphs where uniform sampling at rate $o(\frac{n}{d^2 \ln n})$ results in a vanishingly low probability that the sampled subgraph has a perfect matching. Thus, our sampling rate is tight up to factors of $O(\ln^2 n)$. This lower bound suggests two promising directions for further research: designing an efficiently implementable non-uniform sampling scheme, and designing an algorithm that runs faster than Hopcroft-Karp's algorithm for near-regular bipartite graphs (since the degree of each vertex in the sampled subgraph will be concentrated around the expectation).

## 2 Uniform Sampling for Perfect Matchings: An Upper Bound

In this section, we will establish our main sampling theorem stated below. We will then show in Section 3 that this theorem immediately yields an $O(n^{1.75}\sqrt{\ln n})$ time algorithm for finding a perfect matching in regular bipartite graphs.

**Theorem 2.1** *There exists a constant $c$ such that given a $d$-regular bipartite graph $G(U, V, E)$, a subgraph $G'$ of $G$ generated by sampling the edges in $G$ uniformly at random with probability $p = \frac{cn \ln n}{d^2}$ contains a perfect matching with high probability.*

Our proof is based on a decomposition procedure that partitions the given graph into a vertex-disjoint collection of subgraphs such that (i) the minimum cut in each subgraph is large, and (ii) each subgraph contains $\Omega(d)$ perfect matchings on its vertices. We then show that for a suitable choice of sampling rate, w.h.p. at least one perfect matching survives in each subgraph. The union of these perfect matchings then gives us a perfect matching in the original graph. We emphasize here that the decomposition procedure is merely an artifact for our proof technique. Note that the theorem is trivially true when $d$ is $O(\sqrt{n \ln n})$. So in what follows, we assume that $d$ is $\Omega(\sqrt{n \ln n})$.

### 2.1 Hall's Theorem Witness Sets

Let $G(U, V, E)$ be a bipartite graph. We shall use the following notation. For a graph $G$ and a set of vertices $W$ we denote the number of edges crossing the boundary of $W$ in $G$ by $\delta_G(W)$. Also, we denote the vertex set of $G$ by $V(G)$.

A pair $(A, B)$ with $A \subseteq U$ and $B \subseteq V$ is said to be a *left relevant pair* to Hall's theorem if $|A| > |B|$. Similarly, a pair $(A, B)$ with $A \subseteq U$ and $B \subseteq V$ is said to be a *right relevant pair* to Hall's theorem if $|A| < |B|$.

Given a left relevant pair $(A, B)$, we denote by $E(A, B)$ the set of edges in $E \cap (A \times (V \setminus B))$. Similarly, given a right relevant pair $(A, B)$, we denote by $E(A, B)$ the set of edges in $E \cap ((U \setminus A) \times B)$. We refer to the set $E(A, B)$ as a *witness* edge set if $(A, B)$ is a left or right relevant pair. By Hall's theorem (see, for instance, [4]), to prove Theorem 2.1 it suffices to show that w.h.p. in the sampled graph $G'$, at least one edge is chosen from each witness set. We will focus on a sub-class of relevant pairs, referred to as minimal relevant pairs. A left relevant pair $(A, B)$ is *minimal* if there does not exist another left relevant pair $(A', B')$ with $A' \subset A$ and $E(A', B') \subseteq E(A, B)$. Minimal right relevant pairs are similarly defined. A witness edge set corresponding to a minimal left relevant pair or a minimal right relevant pair is called a *minimal left witness set* or a *minimal right witness set*, respectively. If a graph $G$ has a perfect matching, every minimal witness set must be non-empty. It also follows that any subgraph of $G$ that includes at least one edge from every minimal witness set must have a perfect matching.

A key idea underlying our proof is a mapping from minimal witness sets in $G$ to *distinct* cuts in $G$. In particular, we will map each minimal left witness set $E(A, B)$ to the cut $\delta_G(A \cup B)$. The theorem below shows that this is a one-to-one mapping. The analogous theorem holds for minimal right witness sets.

**Theorem 2.2** *Let $G(U, V, E)$ be a bipartite graph that has at least one perfect matching. If $(A, B)$ and $(A', B')$ are minimal left relevant pairs in $G$ with $E(A, B) \neq E(A', B')$, then $\delta_G(A \cup B) \neq \delta_G(A' \cup B')$.*

**Proof:** Assume by way of contradiction that there exist minimal left relevant pairs $(A, B)$ and $(A', B')$ in $G$ with $E(A, B) \neq E(A', B')$ but $\delta_G(A \cup B) = \delta_G(A' \cup B')$. Then the following conditions must be satisfied for any edge $(u, v) \in E$:

A1. If $u \in (A \setminus A') \cup (A' \setminus A)$ then $v \in (B \setminus B') \cup (B' \setminus B)$. To see this, assume w.l.o.g. that $u \in A \setminus A'$, and then note that if $v \in B \cap B'$, then $(u, v) \in \delta_G(A' \cup B')$ but $(u, v) \notin \delta_G(A \cup B)$. A contradiction. Similarly, if $v \in V \setminus (B \cup B')$, then $(u, v) \in \delta_G(A \cup B)$ but $(u, v) \notin \delta_G(A' \cup B')$. A contradiction.

A2. If $u \in (A \cap A')$ then $v \notin (B \setminus B') \cup (B' \setminus B)$. To see this, consider w.l.o.g. that $v \in (B \setminus B')$. Then $(u, v) \in \delta_G(A' \cup B')$ but $(u, v) \notin \delta_G(A \cup B)$. A contradiction.

In what follows, we slightly abuse the notation and given any (not necessarily relevant) pair $(C, D)$ with $C \subseteq U$ and $D \subseteq V$, we denote by $E(C, D)$ the set of edges in $E \cap (C \times (V \setminus D))$. As an immediate corollary of the properties A1 and A2, we now obtain the following containment results:

B1. $E(A \setminus A', B \setminus B') \subseteq E(A, B)$. This follows directly from property A1 above.

B2. $E(A \cap A', B \cap B') \subseteq E(A, B)$. This follows directly from property A2 above.

We now consider three possible cases based on the relationship between $A$ and $A'$, and establish a contradiction for each case.

**Case 1:** $A \cap A' = \emptyset$. By property A1, if $u \in A \cup A'$ then $v \in B \cup B'$. In other words, there are no edges from $A \cup A'$ to vertices outside $B \cup B'$. Since $|A \cup A'| = |A| + |A'| > |B| + |B'|$, this contradicts our assumption that $G$ has at least one perfect matching.

**Case 2:** $A = A'$. For any edge $(u, v)$ with $u \in A$, property A2 shows that $v \notin (B \setminus B') \cup (B' \setminus B)$. Then $E(A, B) = E(A', B')$. A contradiction.

**Case 3:** $A \cap A' \neq \emptyset$ **and** $A \neq A'$. Assume w.l.o.g. that $A \setminus A' \neq \emptyset$. Since $|A| > |B|$, it must be that either $|A \setminus A'| > |B \setminus B'|$ or $|A \cap A'| > |B \cap B'|$. If $|A \setminus A'| > |B \setminus B'|$, then $(A \setminus A', B \setminus B')$ is a left relevant pair, and by B1, it contradicts the fact that $(A, B)$ is a minimal left relevant pair. If $|A \cap A'| > |B \cap B'|$, then $(A \cap A', B \cap B')$ is a left relevant pair set, and by B2, it contradicts the fact that $(A, B)$ is a minimal left relevant pair. ∎

## 2.2 A Decomposition Procedure

Given a $d$-regular bipartite graph on $n$ vertices, we will first show that it can be partitioned into $k = O(n/d)$ vertex disjoint graphs $G_1(U_1, V_1, E_1), G_2(U_2, V_2, E_2), ..., G_k(U_k, V_k, E_k)$ such that each graph $G_i$ satisfies the following properties:

P1. the size of a minimum cut in $G_i(U_i, V_i, E_i)$ is strictly greater than $\alpha = \frac{d^2}{4n}$.

P2. $|\delta_G(U_i \cup V_i)| \leq d/2$ (hence $G_i$ contains at least $d/2$ edge-disjoint perfect matchings).

The decomposition procedure is as follows. Initialize $H_1 = G$, and set $i = 1$.

1. Find a smallest subset $X_i \subseteq V(H_i)$ such that $|\delta_{H_i}(X_i)| \leq 2\alpha$. If no such set $X_i$ exists, then the decomposition procedure terminates.

2. Define $G_i$ to be the subgraph of $H_i$ induced by the vertices in $X_i$. Also, let $M_i$ denote the number of edges in the cut $\delta_{H_i}(X_i)$.

3. Define $H_{i+1}$ as $H_i$ with vertices from $X_i$ removed.

4. Increment $i$, and go to step (1).

We now prove the following properties of the decomposition procedure.

**Theorem 2.3** *The decomposition procedure outlined above satisfies properties P1 and P2.*

**Proof:** We start by proving that property P1 is satisfied. Suppose that there exists a cut $(C, D)$ in $G_i$ of value less than $\alpha$, i.e. $C \cup D = X_i$ and $\delta_{G_i}(C) = \delta_{G_i}(D) \leq \alpha$. We have $|\delta_{H_i}(C) \setminus \delta_{G_i}(C)| + |\delta_{H_i}(D) \setminus \delta_{G_i}(D)| \leq 2\alpha$ by the choice of $X_i$ in (1). Suppose without loss of generality that $|\delta_{H_i}(C) \setminus \delta_{G_i}(C)| \leq \alpha$. Then $\delta_{H_i}(C) \leq 2\alpha$ and $C \subset X_i$, which contradicts the choice of $X_i$ as the smallest cut of value at most $2\alpha$ in step (1) of the procedure.

It remains to show that $|\delta_G(U_i \cup V_i)| \leq d/2$ for all $i$. In order to establish this property, it suffices to show that $\sum_{i=1}^k M_i \leq d/2$ (recall that $M_i = |\delta_{H_i}(X_i)|$).

We prove the following statements by induction on $k$, the number of decomposition steps:

1. $|U_k \cup V_k| \geq 2d$;
2. $\sum_{i=1}^k M_i \leq d/2$;
3. $k \leq n/d$.

**Base:** $k = 1$ Since $2\alpha = \frac{d^2}{2n} \leq d/2$, we have $M_1 \leq d/2$. It remains to show that $G_1(U_1, V_1, E_1)$ has at least $2d$ vertices. Consider any vertex $u \in U_1$. Let $j \leq d/2$ be the number of edges in $\delta_{H_1}(U_1 \cup V_1)$ that are incident on vertex $u$. Then $u$ must have exactly $(d - j)$ neighbors in $V_1$. Since $|\delta_{H_1}(U_1 \cup V_1)| \leq d/2$, at least one vertex among the neighbors of $u$ in $V_1$ must have all its $d$ neighbors inside $U_1$. Thus $|U_1| \geq d$. Similarly, we can show that $|V_1| \geq d$.

**Inductive step:** $k \to k+1$ Suppose that the $k$-th step has been executed and the algorithm has not terminated yet. Since $k \leq n/d$ by the inductive hypothesis, we have $\sum_{i=1}^{k} M_i \leq (n/d)(2\alpha) = (n/d)\left(\frac{d^2}{2n}\right) \leq d/2$. Consider the cut $(X_k, H_k \setminus X_k)$ of $H_k$. It follows from the previous estimate that $|\delta_{H_k}(X_k)| \geq |\delta_G(X_k)| - d/2$. Hence, we conclude as in the base case that $|X_k| \geq 2d$ and $|H_k \setminus X_k| \geq 2d$. Since at every decomposition step $j \leq k$ at least $2d$ vertices were removed from the graph, we have $k+1 \leq n/d$.

∎

## 2.3 Proof of Theorem 2.1

We now argue that if the graph $G'$ is obtained by uniformly sampling the edges of $G$ with probability $p = \Theta\left(\frac{\ln n}{\alpha}\right)$, then w.h.p. $G'$ contains a perfect matching.

It suffices to show that in each graph $G_i$ obtained in the decomposition procedure, every minimal witness set is hit w.h.p. in the sampled graph (that is, at least one edge in each minimal witness set is chosen in the sampled graph). This ensures that at least one perfect matching survives inside each $G_i$. A union of these perfect matchings then gives us a perfect matching of $G$ in the sampled graph $G'$.

Fix a graph $G_i(U_i, V_i, E_i)$. Let $(A, B)$ be a left or a right relevant pair in $G_i$. Using the fact that our starting graph $G$ is $d$-regular, we get

$$|\delta_G(A \cup B)| \leq 2|E(A,B)| - d.$$

Let $m_A, m_B$ denote the number of edges in $G$ that connect nodes in $A, B$ respectively to nodes outside $G_i$. Then

$$|\delta_{G_i}(A \cup B)| \leq 2|E(A,B)| - d - m_A - m_B.$$

By property P2, since $|\delta_G(U_i \cup V_i)| \leq d/2$, it follows that $|E(A,B) \cap E_i| \geq |E(A,B)| - d/2$. Also, by definition, $|E(A,B) \cap E_i| \geq |E(A,B)| - m_A - m_B$. Combining, we obtain:

$$|\delta_{G_i}(A \cup B)| \leq 2|E(A,B) \cap E_i| - d/2.$$

Thus the set $E(A,B) \cap E_i$ contains at least half as many edges as the the cut $\delta_{G_i}(A \cup B)$. We will now utilize the following sampling result due to Karger [10]:

**Theorem 2.4** [10] *Let $G_i$ be an undirected graph on at most $n$ vertices, and let $\kappa$ be the size of a minimum cut in $G_i$. There exists a positive constant $c$ such that for any $\epsilon \in (0,1)$, if we sample the edges in $G_i$ uniformly with probability at least $p = c\left(\frac{\ln n}{\kappa \epsilon^2}\right)$, then every cut in $G_i$ is preserved to within $(1 \pm \epsilon)$ of its expected value with probability at least $1 - 1/n^{\Omega(1)}$.*

Thus the sampling probability needed to ensure that all cuts are preserved close to their expected value, is inversely related to the size of a minimum cut in the graph. We now show use the theorem above to prove that at least one perfect matching survives in each graph $G_i$ when edges are sampled with probability specified in Theorem 2.1.

By Property P1, we know that the size of a minimum cut in $G_i$ is at least $\alpha = d^2/4n$. Fix an $\epsilon \in (0,1)$. The theorem above implies that if we sample edges in $G_i$ with probability $p = \Theta\left(\frac{\ln n}{\alpha \epsilon^2}\right)$, then for every relevant pair $(A, B)$, w.h.p. the sampled graph contains $(1\pm\epsilon)p|\delta_{G_i}(A\cup B)| = \Omega(\ln n)$ edges from the set $\delta_{G_i}(A \cup B)$.

Note that the set $\delta_{G_i}(A\cup B)$ is not a Hall's theorem witness edge set. However, by Theorem 2.1, we know that for every left (right) minimal witness edge set $E(A, B) \cap E_i$, we can associate a distinct cut, namely $\delta_{G_i}(A \cup B)$, of size at most twice $|E(A, B) \cap E_i|$. We now show that this correspondence can be used to directly adapt Karger's proof of Theorem 2.4 to claim that every witness edge set in $G_i$ is preserved to within $(1 \pm \epsilon)$ of its expected value. We remind the reader that the proof of Karger's theorem is based on an application of union bound over all cuts in the graph. In particular, it is shown that the number of cuts of size at most $\beta$ times the minimum cut size is bounded by $n^{2\beta}$. On the other hand, for the sampling rate given in Theorem 2.4, we can use Chernoff bounds to claim that the probability that a cut of size $\beta$ times the minimum cut deviates by $(1 \pm \epsilon)$ from its expected value is at most $1/n^{\Omega(\beta)}$. The theorem follows by combining these two facts.

Within any piece of the decomposition, let $c_i$ be the number of cuts of size $i$ and let $w_i$ be the number of minimal witness sets of size $i$. We know by the correspondence argument above that every Hall's theorem minimal witness set of size $i$ corresponds to a cut of size at most $2i$, and at most two minimal witness sets (one left and one right) correspond to the same cut.

Now, given a sampling probability $p$, the probability that none of the edges in some minimal witness set are sampled is at most $\sum_i w_i(1-p)^i$, which is at most $\sum_i 2c_i(1-p)^{i/2}$. Therefore the probability that there is no matching in this piece can be at most twice the expression used in Karger's theorem to bound the probability that there exists a cut from which no edge is sampled when the sampling rate is $q$, where $1 - q = (1-p)^{1/2}$, or $p = 2q - q^2$. Hence, it is sufficient to use a sampling rate which is twice that required by Karger's sampling theorem to conclude that a perfect matching survives with probability at least $1 - 1/n^{\Omega(1)}$.

Even though we don't use it in this paper, the following remark is interesting and is worth making explicitly. The remark follows from the additional observation that Karger's proof [10] of theorem 2.4 uses Chernoff bounds for each cut, and these bounds remain the same if we use minimal witness sets which are at least half the size of the corresponding cuts, and then sample with twice the probability.

**Remark 2.5** *There exists a positive constant $c'$ such that for any $\epsilon \in (0,1)$, if we sample the edges in $G$ uniformly with probability at least $p = c'\left(\frac{\ln n}{\alpha\epsilon^2}\right)$, then every minimal witness set in every piece $G_i$ is preserved to within $(1 \pm \epsilon)$ of its expected value with probability at least $1 - 1/n^{\Omega(1)}$. Here $\alpha = d^2/(4n)$, as defined before.*

Putting everything together, the sampled graph $G'$ will have a perfect matching w.h.p. as long as we sample the edges with probability $p > \frac{c \ln n}{\alpha}$ for a sufficiently large constant $c$, thus completing the proof of theorem 2.1. We have made no attempt to optimize the constants in this proof (an upper bound of $\frac{12 \ln n}{\alpha}$ follows from the reasoning above). In fact, in an implementation, we can use geometrically increasing sampling rates until either the sampled graph has a perfect matching, or the sampling rate becomes so large that the expected running time of Hopcroft and Karp [8] algorithm is $\Omega(m)$.

# 3 A Faster Algorithm for Perfect Matchings in Regular Bipartite Graphs

We now show that the sampling theorem from the preceding section can be used to obtain a faster randomized algorithm for finding perfect matchings in $d$-regular bipartite graphs.

**Theorem 3.1** *There exists an $O(\min\{m, \frac{n^{2.5}\ln n}{d}\})$ expected time algorithm for finding a perfect matching in a $d$-regular bipartite graph with $2n$ vertices and $m = nd$ edges.*

**Proof:** Let $G$ be a $d$-regular bipartite graph with $2n$ vertices and $m = nd$ edges. If $d \leq n^{3/4}\sqrt{\ln n}$, we use the $O(m)$ time algorithm of Cole, Ost, and Schirra [6] for finding a perfect matching in a $d$-regular bipartite graph. It is easy to see that $m \leq \frac{n^{2.5}\ln n}{d}$ in this case.

Otherwise, we sample the edges in $G$ at a rate of $p = \frac{cn\ln n}{d^2}$ for some suitably large constant $c$ ($c = 48$ suffices by the reasoning from the previous section), and by Theorem 2.1, the sampled graph $G'$ contains a perfect matching w.h.p. The expected number of edges, say $m'$, in the sampled graph $G'$ is $O(\frac{n^2 \ln n}{d})$. We can now use the algorithm of Hopcroft and Karp [8] to find a maximum matching in the bipartite graph $G'$ in expected time $O(m'\sqrt{n})$. The sampling is then repeated if no perfect matching exists in $G'$. This takes $O(\frac{n^{2.5}\ln n}{d})$ expected running time. Hence, the algorithm takes $O(\min\{m, \frac{n^{2.5}\ln n}{d}\})$ expected time. ∎

Note that by aborting the computation whenever the number of sampled edges is more than twice the expected value, the above algorithm can be easily converted to a Monte-Carlo algorithm with a worst-case running time of $O(\min\{m, \frac{n^{2.5}\ln n}{d}\})$ and a probability of success $= 1 - o(1)$. Finally, it is easy to verify that the stated running time never exceeds $O(n^{1.75}\sqrt{\ln n})$.

# 4 Uniform Sampling for Perfect Matchings: A Lower Bound

We now present a construction that shows that the uniform sampling rate of Theorem 2.1 is optimal to within a factor of $O(\ln^2 n)$. As before, for any graph $G$ the graph obtained by sampling the edges of $G$ uniformly with probability $p$ is denoted by $G'$.

**Theorem 4.1** *Let $d(n)$ be a non-decreasing positive integer valued function such that for some fixed integer $n_0$, it always satisfies one of the following two conditions for all $n \geq n_0$: (a) $d(n) \leq \sqrt{n/\ln n}$, or (b) $\sqrt{n/\ln n} < d(n) \leq n/\ln n$. Then there exists a family of $d(n)$-regular bipartite graphs $G_n$ with $2n + o(n)$ vertices such that the probability that the graph $G'_n$, obtained by sampling edges of $G_n$ with probability $p$, has a perfect matching goes to zero faster than any inverse polynomial function in $n$ if $p = o(1)$ when $d(n)$ satisfies condition (a) above, and if*

$$p = o\left(\frac{n}{(d(n))^2 \ln n}\right)$$

*when $d(n)$ satisfies condition (b) above.*

**Proof:** Note that the theorem asserts that essentially no sampling can be done when $d(n) \leq \sqrt{n/\ln n}$. We shall omit the dependence on $n$ in $d(n)$ to simplify notation.

Define $H^{(k)} = (U, V, E)$, $0 \leq k \leq d$, as a bipartite graph with $|U| = |V| = d$ such that $k$ vertices in $U$ (respectively $V$) have degree $(d-1)$ and the remaining vertices have degree $d$. We will call the vertices of degree $(d-1)$ *deficient*. Clearly, for any $0 \leq k \leq d$, the graph $H^{(k)}$ exists: starting with a $d$-regular bipartite graph on $2d$ vertices, we can remove an arbitrary subset of $k$ edges that belong to a perfect matching in the graph. In the following construction, we will use copies of $H^{(k)}$ as building blocks to create our final instance. In doing so, only the set of deficient vertices in a copy of $H^{(k)}$ will be connected to (deficient) vertices in other copies in our construction.

We now define a $d$-regular bipartite graph $G_n$. Let $\gamma = \left\lceil \frac{d^2 \ln n}{n} \right\rceil$ (note that $\gamma \leq d$ since $d \leq n/\ln n$). We choose $W = \left\lceil \frac{d}{\gamma} \right\rceil$, $k_j = \gamma$ for $1 \leq j < W$, and $k_W = d - \gamma(W-1) \leq \gamma$. We also define $K(n) = \lceil \ln n \rceil$ if $d(n) \geq \sqrt{n/\ln n}$ and $K(n) = \lceil \frac{n}{d^2} \rceil$ otherwise.

The graph $G_n$ consists of $K(n) \cdot W$ copies of $H^{(k)}$ that we index as $\{H_{i,j}\}_{1 \leq i \leq K(n), 1 \leq j \leq W}$. The subgraph $H_{i,j}$ is a copy of $H^{(k_j)}$, where $k_j$ is as defined above. Note that the sum of the number of deficient vertices over each of the parts of $H_{i,j}$, $1 \leq j \leq W$, equals $d$ for all fixed $i$. Moreover, the number of deficient vertices in $H_{i,j}$ is the same for all $i$ when $j$ is held fixed.

We now introduce two distinguished vertices $u$ and $v$ and add additional edges as follows:

1. For every $1 \leq i < K(n)$ and for every $1 \leq j \leq W$, all deficient vertices in part $V$ of $H_{i,j}$ are matched to the deficient vertices in part $U$ of $H_{i+1,j}$ (that is, we insert an arbitrary matching between these two sets of vertices);

2. All deficient vertices in part $U$ of $H_{1,j}$ for $1 \leq j \leq W$ are connected to $u$;

3. All deficient vertices in part $V$ of $H_{K(n),j}$ for $1 \leq j \leq W$ are connected to $v$.

Essentially, we are connecting the graphs $H_{i,j}$ for fixed $j$ in series via their deficient vertices, and then connecting the left ends of these chains to the distinguished vertex $u$ and the right ends of the chains to the distinguished vertex $v$.

We note that the graph $G_n$ constructed as described above is a $d$-regular bipartite graph with $2dK(n)W + 2 = 2n + o(n)$ vertices.

Consider the sampled graph $G'_n$. Suppose $G'_n$ has a perfect matching $M$. In the matching $M$, if $u$ is matched to a vertex in part $U$ of $H'_{1,j}$ for some $1 \leq j \leq W$, then there must be a vertex in part $V$ of $H'_{1,j}$ that is matched to a vertex in part $U$ of $H'_{2,j}$. Proceeding in the same way, one concludes that for every $i, 1 \leq i < K(n)$ there must be a vertex in part $V$ of $H'_{i,j}$ that is matched to a vertex in part $U$ of $H'_{i+1,j}$. Finally, vertex $v$ must be matched to a vertex in part $V$ of $H'_{K(n),j}$. This implies that the sampled graph $G'_n$ can have a perfect matching only if at least one edge survives in $G'_n$ between every pair of adjacent elements in the sequence below: $u \to H_{1,j} \to H_{2,j} \to \ldots \to H_{K(n)-1,j} \to H_{K(n),j} \to v$.

Now suppose that we sample edges uniformly with probability $p$. It follows from the construction of $G_n$ that for any fixed $j$, the probability that at least one edge survives between every pair of adjacent elements in the sequence $u \to H_{1,j} \to H_{2,j} \to \ldots \to H_{K(n)-1,j} \to H_{K(n),j} \to v$ is equal to

$$\left(1 - (1-p)^{k_j}\right)^{K(n)+1} \leq (pk_j)^{K(n)+1}.$$

Hence, the probability that at least one such path survives in $G'_n$ is at most

$$W \left(p \max_{1 \leq j \leq W} k_j\right)^{K(n)+1}$$

by the union bound.

When $d(n) \leq \sqrt{n/\ln n}$, we have $\gamma = 1$, $W = d$, $k_j = 1$ and $K(n) = \lceil n/d^2 \rceil$. So the bound transforms to
$$Wp^{K(n)+1} = dp^{\lceil n/d^2 \rceil+1}, \tag{1}$$
which goes to zero faster than any inverse polynomial function in $n$ when $p = o(1)$ since $K(n) = \lceil n/d^2 \rceil = \Omega(\ln n)$.

When $d \geq \sqrt{n/\ln n}$, we have $k_j \leq \gamma$ where $\gamma = \left\lceil \frac{d^2 \ln n}{n} \right\rceil$, $W = \left\lceil \frac{d}{\gamma} \right\rceil$ and $K(n) = \lceil \ln n \rceil$. Hence, the bound becomes
$$W(p\gamma)^{K(n)+1} = \left\lceil \frac{d}{\gamma} \right\rceil (p\gamma)^{\lceil \ln n \rceil+1}, \tag{2}$$
which goes to zero faster than any inverse polynomial function in $n$ when $p = o\left(\frac{n}{d^2 \ln n}\right)$. This completes the proof of the theorem. ∎

The construction given in Theorem 4.1 shows that the sampling upper bound for preserving a perfect matching proved in Theorem 2.1 is tight up to a factor of $O(\ln^2 n)$.

## Acknowledgments

We thank Rajat Bhattacharjee for many helpful discussions in the early stages of this work.

## References

[1] G. Aggarwal, R. Motwani, D. Shah, and A. Zhu. Switch scheduling via randomized edge coloring. *FOCS*, 2003.

[2] A. Benczur. Cut structures and randomized algorithms in edge-connectivity problems. *PhD Thesis*, 1997.

[3] G. Birkhoff. Tres observaciones sobre el algebra lineal. *Univ. Nac. Tucumán Rev. Ser. A*, 5:147–151, 1946.

[4] B. Bollobas. *Modern graph theory*. Springer, 1998.

[5] R. Cole and J.E. Hopcroft. On edge coloring bipartite graphs. *SIAM J. Comput.*, 11(3):540–546, 1982.

[6] R. Cole, K. Ost, and S. Schirra. Edge-coloring bipartite multigraphs in $O(E \log D)$ time. *Combinatorica*, 21(1):5–12, 2001.

[7] H.N. Gabow and O. Kariv. Algorithms for edge coloring bipartite graphs and multigraphs. *SIAM J. Comput.*, 11(1):117–129, 1982.

[8] J.E. Hopcroft and R.M. Karp. An $n^{5/2}$ algorithm for maximum matchings in bipartite graphs. *SIAM J. Comput.*, 2(4):225–231, 1973.

[9] D. Karger. Random sampling in cut, flow, and network

design problems. *Proceedings of the twenty-sixth annual ACM symposium on Theory of computing*, pages 648–657, 1994.

[10] D. Karger. Using randomized sparsification to approximate minimum cuts. *Proceedings of the fifth annual ACM-SIAM symposium on Discrete algorithms*, pages 424–432, 1994.

[11] D. König. Uber graphen und ihre anwendung auf determinententheorie und mengenlehre. *Math. Annalen*, 77:453465, 1916.

[12] R. Motwani and P. Raghavan. *Randomized Algorithms*. Cambridge University Press, 1995.

[13] A. Schrijver. Bipartite edge coloring in $O(\Delta m)$ time. *SIAM J. on Comput.*, 28:841846, 1999.

[14] D.A. Spielman and N. Srivastava. Graph sparsification by effective resistances. *STOC*, pages 563–568, 2008.

[15] D.A. Spielman and S.-H. Teng. Nearly-linear time algorithms for graph partitioning, graph sparsification, and solving linear systems. *STOC*, pages 81–90, 2004.

[16] J. von Neumann. A certain zero-sum two-person game equivalent to the optimal assignment problem. *Contributions to the optimal assignment problem to the Theory of Games*, 2:5–12, 1953.